\documentclass{article}
\usepackage{spconf,amsmath,graphicx}
\usepackage{multirow}

\title{An Environmental Feature Representation in I-vector Space for Room Verification and Metadata Estimation}
\name{Desmond Caulley}
\address{Georgia Institute of Technology}
\begin{document}
%
\maketitle
\begin{abstract}
    This paper investigates the application of environmental feature representations for room verification tasks and acoustic metadata estimation. Audio recordings contain both speaker and non-speaker information. We refer to non-speaker-related information, including channel and other environmental factors, as e-vectors. I-vectors, commonly used in speaker identification, are extracted in the total variability space and capture both speaker and channel-environment information without discrimination. Accordingly, e-vectors can be extracted from i-vectors using methods such as linear discriminant analysis. In this paper, we first demonstrate that e-vectors can be successfully applied to room verification tasks with low equal-error rate. Second, we propose two methods for estimating metadata information --- signal-to-noise (SNR) and reverberation (T$_{60}$) --- from these e-vectors. When comparing our system to contemporary global SNR estimation methods, in terms of accuracy, we perform favorably even with low dimensional i-vectors. Lastly we show that room verification tasks can be improved if e-vectors are augmented with the extracted metadata information.
\end{abstract}
\section{Introduction}
\label{sec:intro}

With the growing demand for automatic speaker verification (ASV) technology as a biometric modality, there is a need to develop models that are robust to environmental and channel effects. When audio systems record speech from a distant-talking speaker, environmental factors such as background noise or reverberation can be introduced \cite{Ganapathy2011FeatureNF, Gammal2005CombatingRI}. ASV accuracy is severely degraded by these extraneous factors \cite{zhao2014robust, shabtai2008effect, castellano1996speaker}. Capturing and analyzing these factors, therefore, is an important first step in the development of robust ASV.

I-vectors have been the primary force behind the advances in speaker verification and identification tasks \cite{Dehak09-SVM,Dehak2011FrontEndFA}. The traditional i-vector pipeline includes a universal background model (UBM) and a large projection matrix, commonly known as the T-matrix, which is trained using a maximum-likelihood estimation framework. The T-matrix  maps Gaussian mixture model (GMM) supervectors from the UBM to a lower dimensional space. The output of this transformation is known as an i-vector. Last in this pipeline is a probabilistic linear discriminant analysis (PLDA) classifier \cite{Ioffe2006ProbabilisticLD}. PLDA is typically used to compare two i-vectors to check the likelihood that they come from the same versus different speakers \cite{Kenny2010BayesianSV, GarciaRomero2011AnalysisOI}.

Despite the success of i-vectors in the speaker verification domain, there has been little research in their use for environment/channel related tasks. I-vectors are extracted in the total variability space in the sense that they make no distinction between channel factors and speaker factors. Just as speaker information is contained in i-vectors, so is channel-environment information.

For this research, we extract environment information from i-vectors via linear discriminant analysis (LDA). The resultant embeddings are known as LDA e-vectors. In 2017, Feng, {\em et al.}, appended LDA e-vectors to original speech features as an adaptive bias in their work on noise condition classification  \cite{Feng17-COP}. They showed that system accuracy is significantly improved with this feature adaptation. 


The work by Feng, {\em et al.} considers limited noise types and noise variations. Overall, the network they trained can be seen as having  a total of 60 noise conditions. While the present study is related to Feng's work in e-vector extraction method, it further generalizes to the case where there might be thousands of noise conditions. Feng's work examined estimating factors such as car speed (3 different speeds), air conditioning status (on/off), and other similar information from i-vectors. Our work here combines thousands of noise types, instead of from a limited set, to create  virtual rooms. Specifically, we show that e-vectors contain environment information that can also be used for room verification (RV). Most significantly, these e-vectors can be extracted in a low-dimension subspace, effectively easing back-end computations.

Furthermore, this research explores a new idea of e-vectors as a toolkit that contains valuable metadata information. We  propose estimating two different acoustic metadata values---signal-to-noise (SNR) and reverberation (T$_{60}$)---from the e-vectors. We designed two different models for this task. One is a traditional ridge regression and the other is a bottleneck neural network (BN-NN) styled regression model. We show that augmenting i-vectors with estimated metadata information can further improve accuracy.

\subsection{E-vector Extraction via LDA}

LDA is a commonly employed technique in statistical machine learning that aims to find a linear combination of features that will maximally discriminate between classes. In a multi-class setting, LDA functions as a dimension-reducing transformation that minimizes the scatter within each class while maximizing the scatter between classes.
However, this transformation makes some assumptions about the input features. First, that independent variables are normally distributed for each class. Second, that Gaussian models for each class are assumed to share the same covariance matrix.

In speaker identification and verification tasks, LDA serves as a channel compensation technique to increase the inter-speaker variability while minimizing the intra-speaker variability (due to channel variability). 

For room verification tasks, a similar solution can be formulated. Instead of channel compensation, LDA can be used as a speaker compensation technique to minimize the effect of intra-room variability. The between and within covariance matrices are given by
\begin{equation}
    S_b = \frac{1}{R} \sum_{r=1}^{R} (\overline{i}_r - \overline{i})(\overline{i}_r - \overline{i})^T ~and~
\end{equation}
\begin{equation}
    S_w = \frac{1}{R} \sum_{r=1}^{R} \frac{1}{n_r} \sum_{k=1}^{n_r} (i_{r,k} - \overline{i}_r)(i_{r,k} - \overline{i}_r)^T
\end{equation}
where $R$ is the total number of rooms, $\overline{i}$ is the mean of all i-vectors across all rooms, $\overline{i}_r$ is the mean of i-vectors belonging to room $r$,  $i_{r,k}$ is the $k^{th}$ utterance from room $r$, and $n_r$ denotes the total number of utterances belonging to room r.

The core of LDA involves maximizing the Rayleigh coefficient. The outcome of this optimization is a projection matrix $A_{LDA}$ that is comprised of the top $j$ eigenvectors corresponding to the $j$ largest eigenvalues ($j$ is the rank of the $A_{LDA}$ matrix). Specifically, the solution to the following generalized eigenvalue problem is the projection obtained by the optimization mentioned above:
\begin{equation}
    S_bv = \Lambda S_w v
\end{equation}
Here, $\Lambda$ is a diagonal matrix of eigenvalues. The projection matrix and the transformed feature vector are given by the following equations:
\begin{equation}
    A_{LDA} = [v_1 ..... v_j] 
\end{equation}
\begin{equation}
    \Phi_{_{LDA}} (i) = A^T_{_{LDA}}i
\end{equation}

\section{DATASET CREATION}
\label{sec:pagestyle}
We constructed a dataset of various ``virtual rooms'' for experimentation. To accomplish this, we used three different corpora --- VoiceHome \cite{Bertin-FRE}, MUSAN \cite{musan2015}, and TIMIT \cite{Lopes2012TIMITAC}. The VoiceHome corpus contains recordings of room impulse responses in various acoustic settings. The room impulse recordings are typically less than 0.5 seconds. Next is the MUSAN corpus, which contains various speech, music, and noise recordings. Last is the TIMIT dataset. It is a balanced collection of clean speech recordings from various speakers. All audio in the corpora was sampled at a rate of 16 kHz.

A virtual room is defined as a combination of noise, music, and room impulse response. The speech data from the TIMIT corpus is added to introduce speaker variability. Just as speech from a particular speaker might have variable noise conditions in the background for speaker verification task, we wanted our rooms to have speech as the ``noise'' in the background.

The first step in creating a room is to estimate the reverberation time, T$_{30}$, of the room impulse response recordings. To do this, we first normalize the audio and check the time it takes for the signal's energy to decay by 30 dB. We multiply this value by 2 to obtain our T$_{60}$. The next step is to then modify the T$_{60}$ value of that recording. This can be accomplished by raising the original impulse recording to the power $\alpha$ where $\alpha = \frac{T_{60} estimated}{T_{60} desired} $. This procedure was significant since it allowed us to control the reverberation level in each virtual room manually.

Many different room types were created. One room type is a ``complete\_room''or one with a stationary, nonstationary, and music mixed in equal ratios. We then combined this audio with a reverberant speech at a specific SNR. By appropriately weighting the noise conditions created before mixing with the clean speech, we can control the SNR value. The reverberant speech is created by convolving the modified room impulse recording with a randomly selected speech in the TIMIT corpus. Other room types were created. These room types were essentially complete rooms with one or more factors missing. For example, a ``no\_nonstat\_noise" is a complete room recording without the added nonstationary noise and ``music\_rir" is room with only room impulse mixed with music.
The SNR values used in room creation ranges from 5 dB to 25 dB. The T$_{60}$ values, on the other hand, range from 50 milliseconds to 500 milliseconds. By analyzing the persistent properties of a given audio signal, e-vectors can effectively determine the deviations from the persistent properties and capture that as noise conditions.

For training, we created 3500 different complete rooms. Each complete room had 50 instances, where the difference between the instances is the speaker variabilities. On average, each audio in the training set is about 16 seconds long. For validation, we created another set of 500 complete rooms. Last was the enrollment/testing set. We created 500 different rooms of each of the 5 room types. Each room had 4 testing audio instances. Also, each test audio was roughly 10 seconds long.
\section{Experiments}
\label{sec:experiments}
To demonstrate that e-vectors contain valuable non-speaker-related information, we set up a room verification task using the synthetic data described above. Here we use equal error rate (EER) as a metric for evaluating this feature representation's viability. After that, we set up a series of experiments to extract metadata information from the e-vectors. We proposed two methods for this extraction and compared the results to a baseline of Waveform Amplitude Distribution Analysis (WADA) SNR predictor \cite{kim2008robust}. The WADA algorithm assumes that clean speech has Gamma distribution with parameter  0.4  and estimates  SNR  by examining the amplitude distribution of audio and how it deviates from this Gamma distribution. The metric for evaluating metadata extraction performance is mean absolute error (MAE).

\subsection{Room Verification}
\label{ssec:roomverification}
We set up a room verification task similar to traditional speaker verification for this experiment. However, instead of using speakers as labels, we used the various rooms we created as the labels. The same 3500 complete rooms were used as the development set across all experiments. We had a separate dataset for each of the five-room types for testing. We enrolled 500 separate rooms for each room type and tested on those rooms.

The room verification experiments can be seen as a proxy task to show that i-vectors indeed carry non-speaker-related information, which can be further extracted using LDA with the different rooms as labels. The size of the e-vectors is directly related to the choice of LDA dimensionality. The experiments examined the relationship between the e-vector dimensionality and its discriminative power.

\subsection{Metadata Estimation Using Linear Regression}
\label{ssec:metadataestimation}
The next series of experiments focused on estimating SNR and T$_{60}$ from a given e-vector. The data used during i-vector training were not only labeled according to the room with which they belong but also the SNR and T$_{60}$ values used during creation. Thus e-vectors, extracted from i-vectors, had a T$_{60}$ and SNR label attached to them. Here we trained two different models that can be used to predict the metadata value given an e-vector. The models were optimized over a validation set and tested on the same test set used in the room verification task.

\begin{figure}[tb]
  \centering
  \includegraphics[scale=0.5]{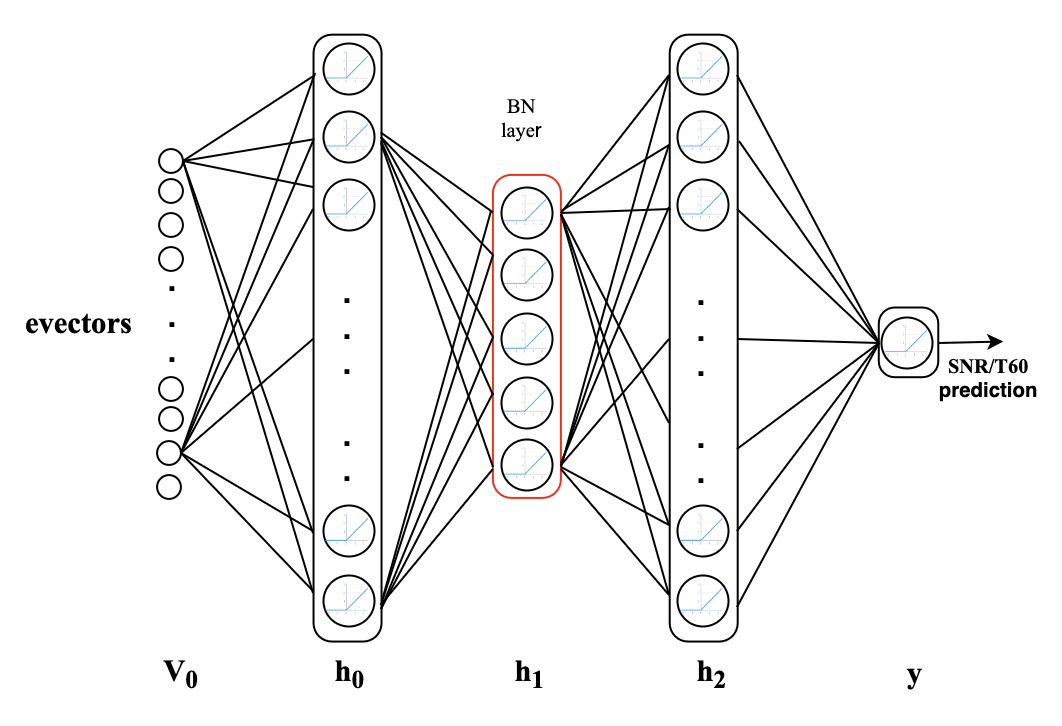}
  \caption{Bottleneck Neural Network for T$_{60}$/SNR Estimation.}
  \label{fig:evec_bnn}
\end{figure}

\subsection{Augmenting E-vectors with Metadata}
\label{ssec:subsubhead}
The last series of experiments examined whether the room verification task could be improved if i-vectors were augmented with extracted metadata. Here, we trained a new LDA and PLDA, which considers the metadata augmentation.

\section{Proposed Models}
\label{sec:models}

We propose two methods for estimating metadata from e-vectors

\subsection{Ridge Regression}
\label{ssec:ridgeregression}
Ridge regression (RR), also known as Tikhonov Regularization, is an instance of linear regression (LR) where a penalty term is added to the standard LR formulation to introduce some bias. RR can be summarized using the objective function below:
\begin{equation}
    \hat{\beta} = \left| \arg\min_{\beta} \lVert y - X\beta \rVert + \lambda \lVert \beta \rVert \right|^2 \\
\end{equation}
We can obtain standard LR by setting the lambda to zero in the equation above. The solution of the ridge regression can be summarized by the equation below: \\
\begin{equation}
    \hat{\beta} = (X^T X + \lambda I)^{-1}X^T y \\
\end{equation}

\subsection{Bottleneck Neural Network}
\label{ssec:metadata_ivectors}
We constructed an architecture with a BN-NN type topology to make T$_{60}$ and SNR predictions when given e-vectors as input. BN-NN is a particular DNN architecture where the dimensionality of one of the hidden layers is significantly lower than those of the surrounding layers. For our model, we have three hidden layers. The first and third layers have an equal number of nodes, while the dimensionality of the second layer is significantly lower than the other two. 

The configuration is $M\times 20 \times 5 \times 20 \times 1$, where $M$ refers to the size of the input e-vector extracted using LDA. The output is a single number that represents the estimated SNR or T$_{60}$ value. We decided on this configuration for several reasons. First, increasing the dimensionality of the layers did not improve the metadata estimation. Second, it increased computational complexity along with the number of parameters to estimate. The activation function used in the entirety of this network is the rectified linear unit  (ReLU).

\section{Results}
\label{sec:results}
\begin{figure}[tb]
  \centering
  \includegraphics[scale=0.6]{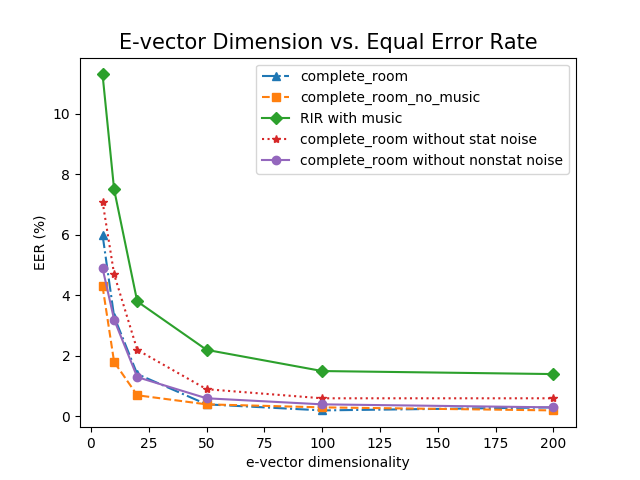}
  \caption{Room Verification Error Rate in different room types.}
  \label{fig:evec_vs_eer}
\end{figure}

Despite experimentation on synthetic data, we can draw several meaningful conclusions about the nature of environment information in i-vector space. \ref{fig:evec_vs_eer} shows that across all room types, we can achieve EER comparable to that of traditional speaker verification systems. With dimensionality set at 50, we obtained an EER of less than 2.5\% on all room types. Even though we trained using the ``complete room" dataset, testing on unseen room types outperformed testing data from even the complete room data in some cases. This shows that our trained model generalizes well to new environmental conditions.

One key outcome of the experiment is that e-vectors can be extracted at relatively low dimensionality. Figure \ref{fig:evec_vs_eer} shows that EER decreases exponentially as e-vector dimensionality increases.  However, it would be optimal, particularly for backend computations, if this feature embedding lies in a lower-dimensional subspace. Going from 5-dim to 200-dim, the EER value decreases by 14.5 times. However, it only decreases by 2.7, going from 20-dim to 200-dim. This means that at an evector-dimension of 20; we achieve about 78\% of gain in EER value decline, showing that most environment information is contained at a relatively low dimension.
\subsection{Metadata Estimation Results}
\label{sec:experiments}
\begin{figure}[tb]
  \centering
  \includegraphics[width=\linewidth]{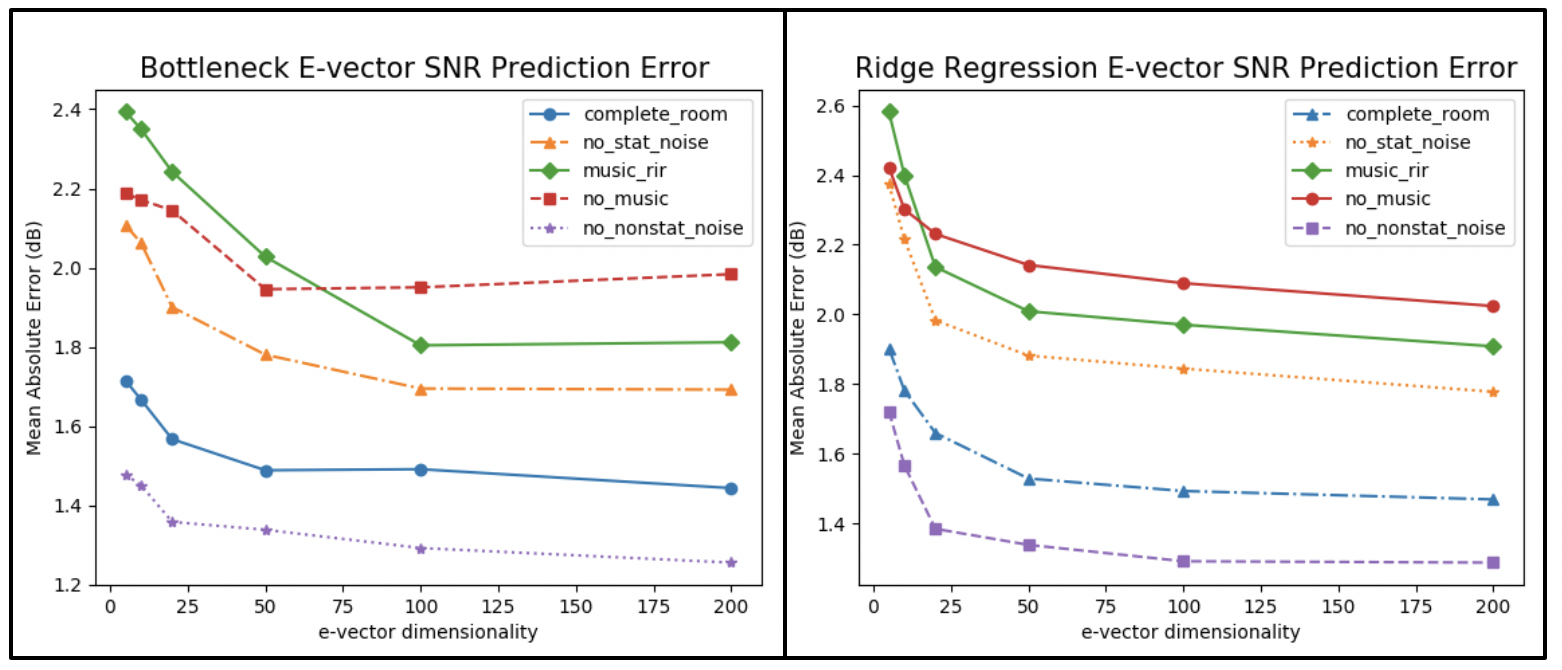}
  \caption{SNR Prediction on Proposed Methods.}
  \label{fig:snr_results}
\end{figure}

\begin{figure}[tb]
  \centering
  \includegraphics[width=\linewidth]{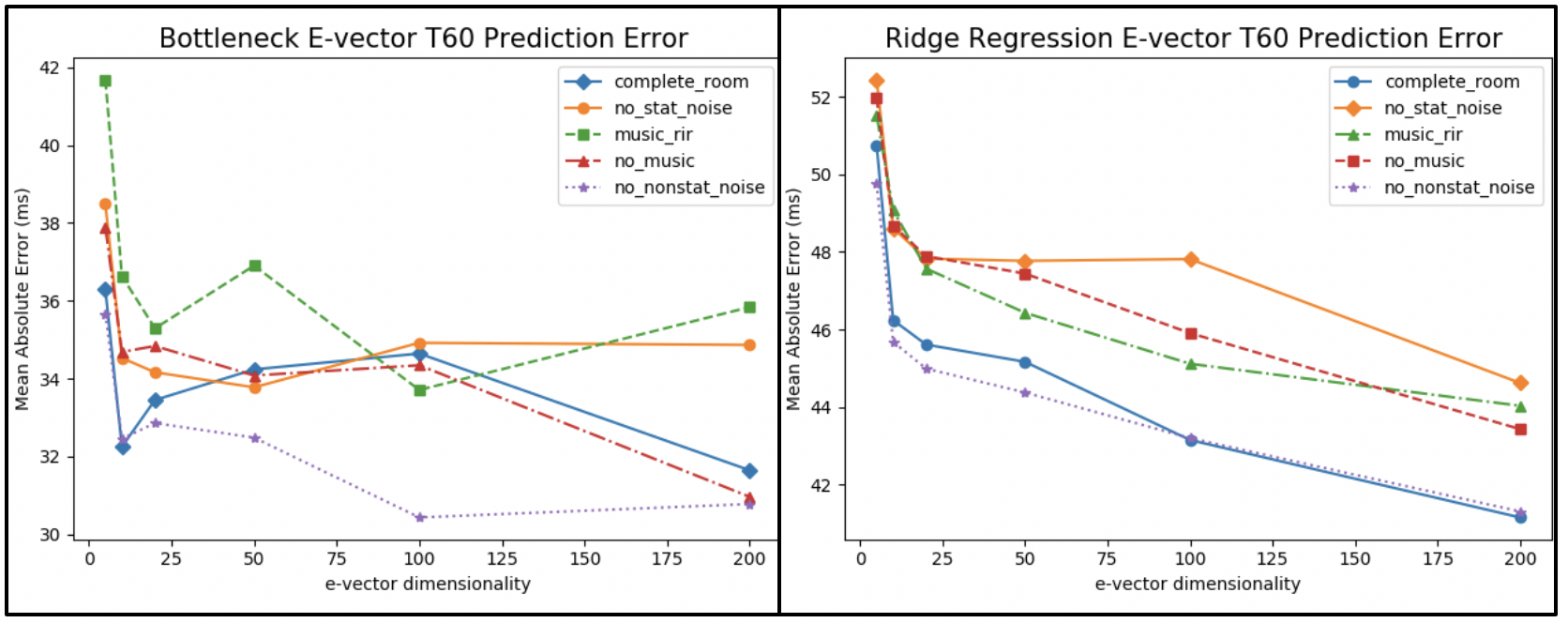}
  \caption{T$_{60}$ Estimation on Proposed Methods.}
  \label{fig:t60_results}
\end{figure}

We examined SNR predictions using e-vectors with mean absolute error (MAE) as the metric. Figure \ref{fig:snr_results} highlights the outcome of this experiment. For both the bottleneck and ridge regression methods, SNR prediction decreases as e-vector dimensionality decreases as expected. SNR prediction using the bottleneck architecture outperforms the ridge regression across all rooms types and e-vector dimensionality. On average, predictions using bottleneck led to a 6\% reduction in prediction error across all room types. Considering a 50-dimensional e-vector as an example, SNR prediction error using ridge regression was 1.78 compared to 1.72 for the bottleneck.

\begin{table}[!ht]
    \centering
    \begin{tabular}{  |c || c |c | c |}
    \hline
                          & \bf{WADA} & \bf{RR} & \bf{BN-NN}   \\ 
    \hline
    complete\_room                       & 2.42 & 1.52 & \bf{1.49}   \\ 
    \hline
    no\_music              & 3.89 & 2.14 & \bf{2.05} \\ 
    \hline
    music\_rir                              & 3.47 & 2.01 & \bf{1.99}   \\ 
    \hline
    no\_stat\_noise                 & 3.52 & 1.88 & \bf{1.80}   \\ 
    \hline
    no\_nonstat\_noise                & 2.39 & 1.34  & \bf{1.29}  \\ 
    \hline
    \end{tabular}
    \caption{SNR Mean Absolute Error in Different Room Types (dB). Ridge Regression (RR); Bottleneck Neural Network (BN-NN) }
    \label{tab:EER_VR}
\end{table}

The proposed models consistently outperformed the WADA SNR prediction, as shown in Table 1. The average MAE using the WADA method is roughly 3.14 but only 1.72 using the bottleneck neural network model. This is an 82\% decrease in prediction error. The experiment shows that our proposed method for SNR estimation is resilient to diverse noise types considering we created a dataset using thousands of real-life noises.

For comparison, we saw the same behavior for the T$_{60}$ estimation proposed bottleneck approach outperformed the ridge regression by reducing prediction error of about 27\% on average. The average prediction error for bottleneck-based T$_{60}$ estimation is 34 milliseconds. Unexpectedly, however, prediction error did not consistently decrease as e-vector dimensionality increases in the neural network solution.

\subsection{Augmentation}
\label{sec:augmentation}
Finally, we augmented i-vectors with estimated metadata before LDA and PLDA training. Across all room types, adding this extra bit of information about the room either left the error rate unchanged or, in some cases, significantly reduced an error rate. Table 2 below shows that by augmenting i-vectors with both SNR and T$_{60}$, the error rate was reduced by 16.9\%. 
\begin{table}[!ht]
    \centering
    \begin{tabular}{  |c || c |c | c | c  |}
    \hline
     & \multirow{3}{*}{\bf{i-vec}} & \bf{i-vec} & \bf{i-vec} & \bf{i-vec } \\
     & & \bf{\& SNR} & \bf{\& T60} & \bf{\& SNR} \\
     & & & &  \bf{\& T60} \\
    \hline
    complete\_room                       & 3.3 & 3.0 & 3.0  & \bf{3.0} \\ 
    \hline
    no\_music               & 1.8 & 1.5 & 1.7 & \bf{1.7}\\ 
    \hline
    music\_rir                              & 7.5 & 7.0 & 7.2 & \bf{6.7} \\ 
    \hline
    no\_stat\_noise                 & 4.7 & 4.3 & 4.4 & \bf{3.8}  \\ 
    \hline
    no\_nonstat\_noise               & 3.2 & 2.6  & 2.5 & \bf{2.4} \\ 
    \hline
    \end{tabular}
    \caption{EER $(\%)$ after i-vector augmentation with metadata}
    \label{tab:EER_VR}
\end{table}

\section{Conclusions}
\label{sec:conclusions}
The research described shows i-vectors contain valuable non-speaker related/environment information called e-vectors. These embeddings can be used to estimate acoustic information like SNR and T${60}$. We proposed two models including a bottleneck neural network model that can estimate SNR with error margin of roughly 1.72 dB using 50-dim e-vector. Lastly, we show that augmenting i-vectors with metadata information can improve room verification performance.  
Future work may focus towards exploring how we can improve speaker diarization using information from e-vectors. For example, applying room specific model to diarize an audio recording once an e-vector gives information about the acoustic nature of the room the audio was recorded.

\vfill\pagebreak

\bibliographystyle{IEEEbib}
\bibliography{refs}

\end{document}